# An attempted replication of Hackl, Koster-Hale, Varvoutis (2012)

Edward Gibson[a], Roger Levy[a,b]

[a]Department of Brain and Cognitive Sciences, MIT
[b]Department of Linguistics, University of California San Diego

## Abstract

Hackl, Koster-Hale & Varvoutis (2012; Journal of Semantics, 29, 145–206; HKV) provide data that suggested that in a null context, antecedent-contained-deletion (ACD) relative clause structures modifying a quantified object noun phrase are easier to process than those modifying a definite object NP. HKV argue that this pattern of results supports a quantifier-raising (QR) analysis of both ACD structures and quantified NPs in object position: under the account that they advocate, both ACD resolution and quantified NPs in object position require movement of the object NP to a higher syntactic position. The processing advantage for quantified object NPs in ACD is hypothesized to derive from the fact that – at the point where ACD resolution must take place – the quantified NP has already undergone QR whereas this is not the case for definite NPs. Here, we report attempted replications of their self-paced reading Experiments 1 and 2. We do not replicate the critical interactions in any of the words immediately following the disambiguating verb in either experiment. Putting these observations together with the observation that it was only post-hoc analysis decisions that were responsible for HKV's observed effects in the first place (Gibson et al., submitted), we conclude that the experiments reported by HKV should not be viewed as providing evidence for the ACD quantifier raising processing effect.

## 1. Introduction

In a recent attempt to find on-line evidence for the existence of Quantifier Raising (QR), Hackl, Koster-Hale & Varvoutis (HKV; 2012) reported that antecedent-contained-deletion (ACD) relative clauses modifying an object noun phrase (NP) whose determiner is "every" (e.g., "every doctor... was", as in (1a)) were easier to process than those modifying a object NP whose determiner is "the" (e.g., "the doctor... was"). In contrast, in full verb controls (as in (1b)), "every" had no advantage over "the".

(1) The understaffed general hospital was negotiating with every / the doctor ...
     a. that the nonprofit medical organization was...   [ellipsis condition]

      b. that the nonprofit medical organization funded...   [full verb condition]

... in order to arrange for free vaccination clinics.

Critically, HKV observed an interaction in the reading times such that the "the" / ellipsis condition was especially slow in the region soon after the elided verb was processed, compared to the other three conditions.  HKV argued that this pattern of results supports a QR analysis of both ACD structures and quantified NPs in object position. Under this account, both ACD resolution structures and NPs with "every" in object position require movement of the object NP to a higher syntactic position. The advantage for object NPs whose determiner is "every" relative to object NPs whose determiner is "the" in ACD structures was hypothesized to arise from the fact that the quantified NP, but not the definite NP, has already undergone QR by the time ACD has to be resolved. The two control full verb conditions require no movement at the verb, and are therefore also processed relatively quickly.

HKV reported results from an off-line acceptability rating study and two self-paced reading studies in support of their account.  More recently, Gibson et al. (2014) showed that HKV's experimental design was confounded, with the consequence that their results could be explained by pragmatic and plausibility factors. But unlike the QR hypothesis, the pragmatic hypothesis doesn't predict that the effects should occur right at or immediately after the elided verb in question. In order to better understand the on-line data in HKV's self-paced reading experiments, we therefore attempted to replicate HKV's self-paced reading experiments.

## 2. Attempted replication of HKV Experiment 1

### 2.1 Procedure

We used in-house self-paced reading software written by Harry Tily to run HKV's Experiment 1 on Amazon.com's Mechanical Turk.  Using this software, sentence materials were presented one word at a time, and reaction times between button presses were recorded.  There was a comprehension question following each trial, as in HKV's self-paced reading experiments.

### 2.2 Materials

The target materials were slightly modified versions of 60 HKV's Experimental 1 items. At the time that we ran this replication attempt in 2012, we only had access to HKV's target materials, in an appendix in their paper, but we did not have access to HKV's filler materials or to the comprehension questions for the target materials.  We therefore constructed our own set of 60 filler items, with questions, and comprehension questions for the target materials

Each of the target items consisted of an initial subject noun phrase of 4-5 words long, the auxiliary verb "was", a progressive verb, sometimes a particle, then the ACD relative clause, and the following critical region for analysis. An example is provided in (1):

(1) The understaffed general hospital was negotiating_with the/every doctor that the nonprofit medical organization was/funded in order to arrange for free vaccination clinics.

In (1), the initial subject noun phrase is "The understaffed general hospital". The main verb is "negotiating", and it is followed by the particle "with" in this item. (We present these two words with an underscore in (1) above to indicate that these two words were presented together in the same region. The underscore was not presented to the participants, however.) The ACD relative clause then consisted of "the/every doctor that the nonprofit medical organization was/funded" in this item. The critical region then consisted of the four words that follower, word presentation regions 15-18 in (1), "in order to arrange".

We modified these items so that the initial NP was always 4 words (sometimes deleting or editing a word in the original HKV items), and the progressive verb region contained the following particle, if there was one (46 of HKV's items included particles; the remaining 14 did not). This way, all items consisted of a subject noun phrase consisting of four words, then the auxiliary verb "was", then a progressive verb (which was presented with a particle for some verbs), then the ACD relative clause, and the following critical regions. Items 1-15, 17-27, 35-37, 39-43, 45-55 were edited so the verb and following particle were presented together (e.g., "negotiating with" were presented together as one region). (The item numbers correspond to the numbers in the appendix of HKV's paper.)

Other item-by-item changes are noted below. These changes were made only in order to regularize the regions in the initial noun phrase, and to correct a few confusing word choices in the original items. Note that the ACD region and the following critical regions were unchanged from HKV's original materials.

Items 1-23, 25, 27, 28, 35-37, 39-43, 45-55: no further changes

24: "the absent minded assistant" -> "the absent-minded new assistant"

26: "the foolish bus boy" -> "the foolish Italian busboy" ("busboy" is usually spelled as a single word)

29: "The young and trendy stylist" -> "The young and_trendy stylist" ("and trendy" were presented as a single region in order to keep the initial subject NP at 4 words)

30: "The overly competitive young acrobat" -> "overly" was deleted

31: "The exceptionally knowledgeable history teacher" -> "exceptionally" was deleted

32: "The young but respected lawyer" -> "but" was deleted

33: "The very passionate young teacher" -> "very" was deleted

34: "The very busy international banker" -> "very" was deleted

38: "The rather clever head detective" -> "rather" was deleted

44: "photo copying"" -> "photo-copying"

56: "The terribly stubborn soccer referee" -> "soccer" was deleted

57: "The newly hired geography tutor" -> "The new geography tutor"

58: "The newly hired professional valet" -> "The extremely professional valet"

59: "The usually grumpy postal worker" -> "usually" was deleted

60: "The newly trained young nanny" -> "The newly_trained young nanny"

We reiterate that these changes simply regularized the form of the initial matrix NP for the sentence, and the matrix verbal region. Both of these regions occurred well before the critical region, which was not modified from HKV's items.

## 2.3 Participants

78 participants from Amazon.com's Mechanical Turk took part in the self-paced reading study. We analyzed only participants who identified themselves as native speakers of English from the US. This resulted in omitting the data from 5 of the initial 78 participants. One additional participant was removed for answering near chance on the comprehension questions (49% accuracy). All other participants answered 77% or more of the comprehension questions correctly. 72 participants remained for analysis.

## 2.4 Results

Accuracy data. Participants answered 89.2% of the comprehension questions correctly, across target and filler materials, and 86.2% of the questions following only target items. The range of accuracies on particular target items varied from 46% to 98% (with one mistakenly coded question, at 10%). All items except for the one at 46% were answered at 60% or better. We present analyses including items answered at greater than 60% accuracy. Analyses including all items result in very similar statistical patterns.

Reading time data. In order to make these analyses as comparable as possible with HKV's experiments, we analyzed the RT data in the same way that HKV did. Other ways of analyzing the data result in similar statistical results, with all statistical tests on the same side of significance, no matter which analysis we examined. Full analyses are available in our R analysis file (available at https://osf.io/t6anw ).

We first omitted RTs below 90 msec and greater than 3 seconds. A regression equation was then computed for each participant, predicting reading time from word length using all trials. At each word position, the reading time predicted by the participant's regression equation was subtracted from the measured reading time to obtain a residual reading time. For each participant, residual reading times beyond two standard deviations from the mean for a given condition and position were excluded from analyses.

Following HKV, we focus our analysis on the four words following the disambiguating verb in the ACD relative clause: word regions 15-18 ("in order to arrange" in (1)). HKV found an interaction between the verb type and determiner at the third word in this region. We present means and 95% confidence intervals for each of the target regions in the figures below. Figure 1 presents the residual RTs across the sentences. For comparison purposes, we also present HKV's Figure 1, on their residual RTs in their Experiment 1 in our Figure 2. Similar RT patterns are visible in our data as in HKV's. Figure 3 presents the residual RTs for our experiment in word regions 15-18. Then we present raw RTs across the sentences in Figure 4, and broken down by region in Figure 5.

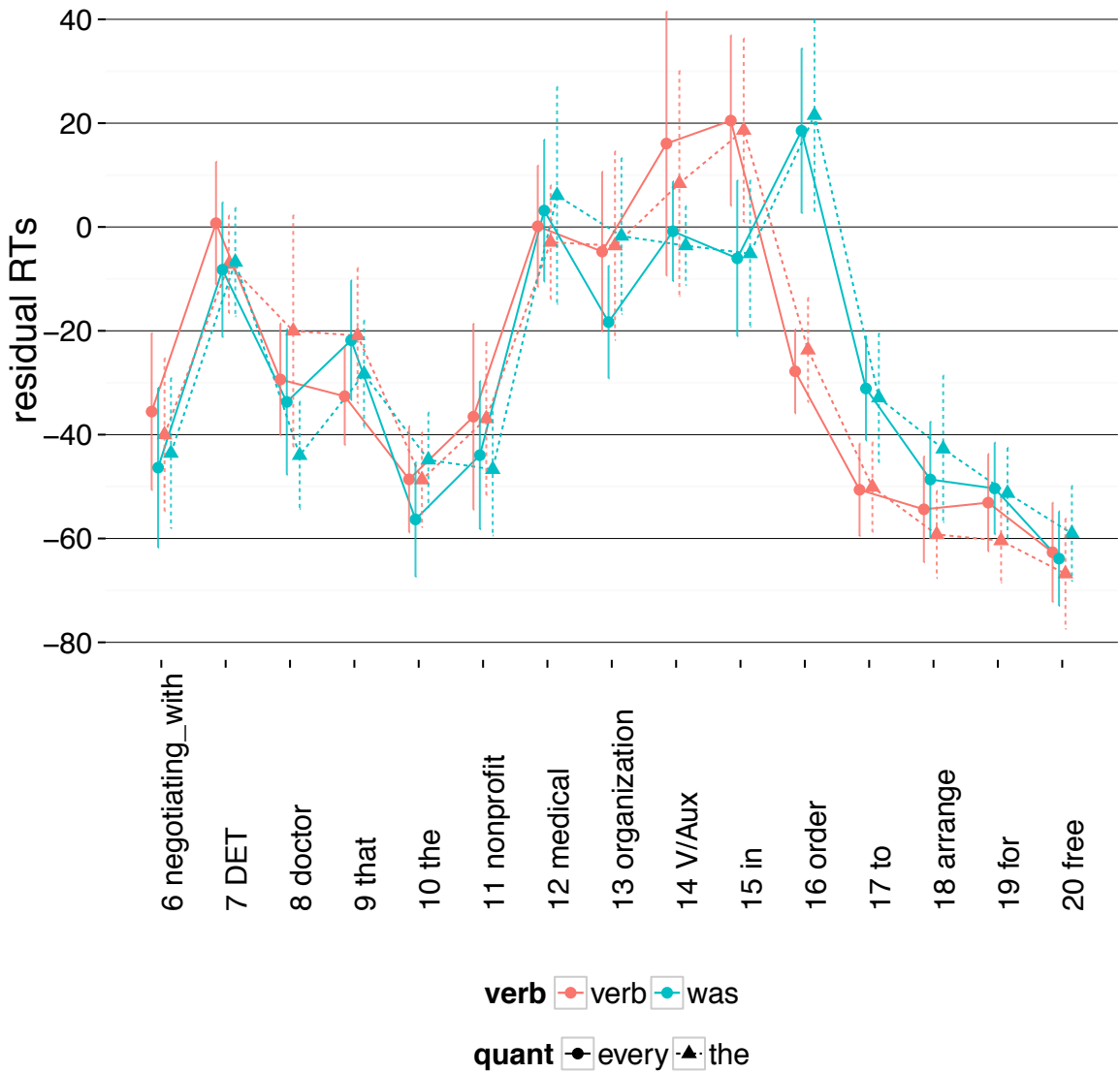

Figure 1: Word-by-word residual reading times for Replication of HKV Experiment 1. Error bars are 95% confidence intervals on participant means.

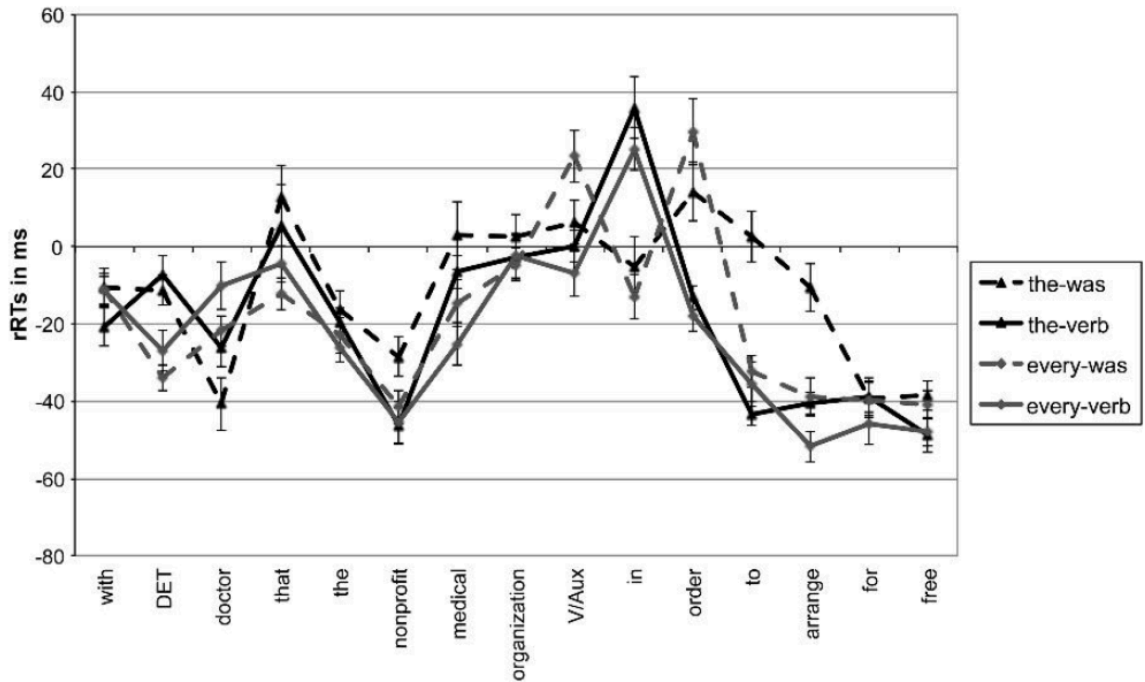

**Figure 1** Average residual RTs for region of interest for Experiment 1 ($n = 44$).

Figure 2: Figure 1 from HKV: Average residual RTs for Experiment 1. Note that the error bars here are smaller than in our plots since they represent standard errors, as opposed to 95% CIs.

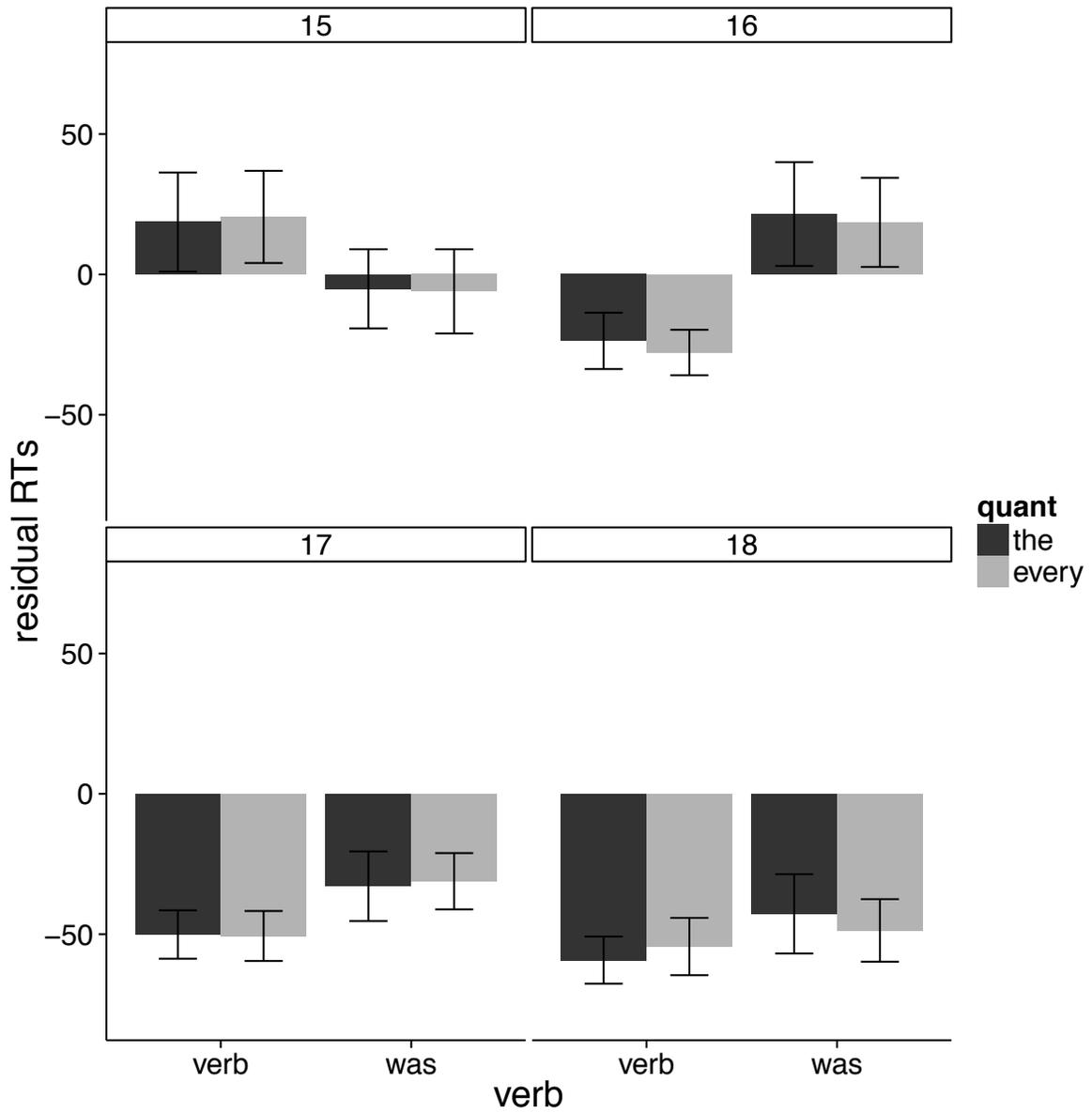

Figure 3: Residual reading times for HKV critical region in Replication of Experiment 1. Error bars are 95% confidence intervals on participant means.

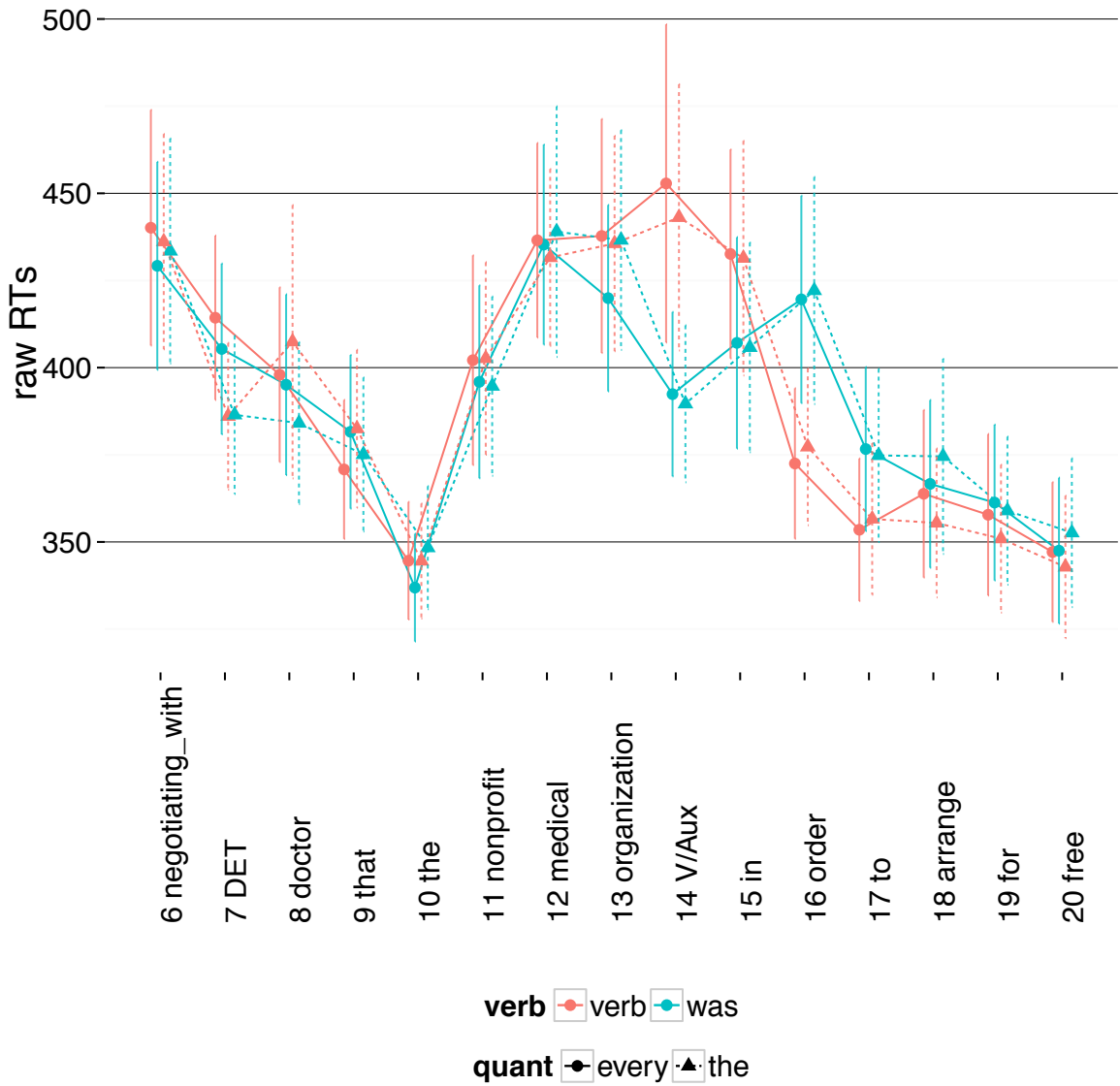

Figure 4: Word-by-word raw reading times for Replication of HKV Experiment 1. Error bars are 95% confidence intervals on participant means.

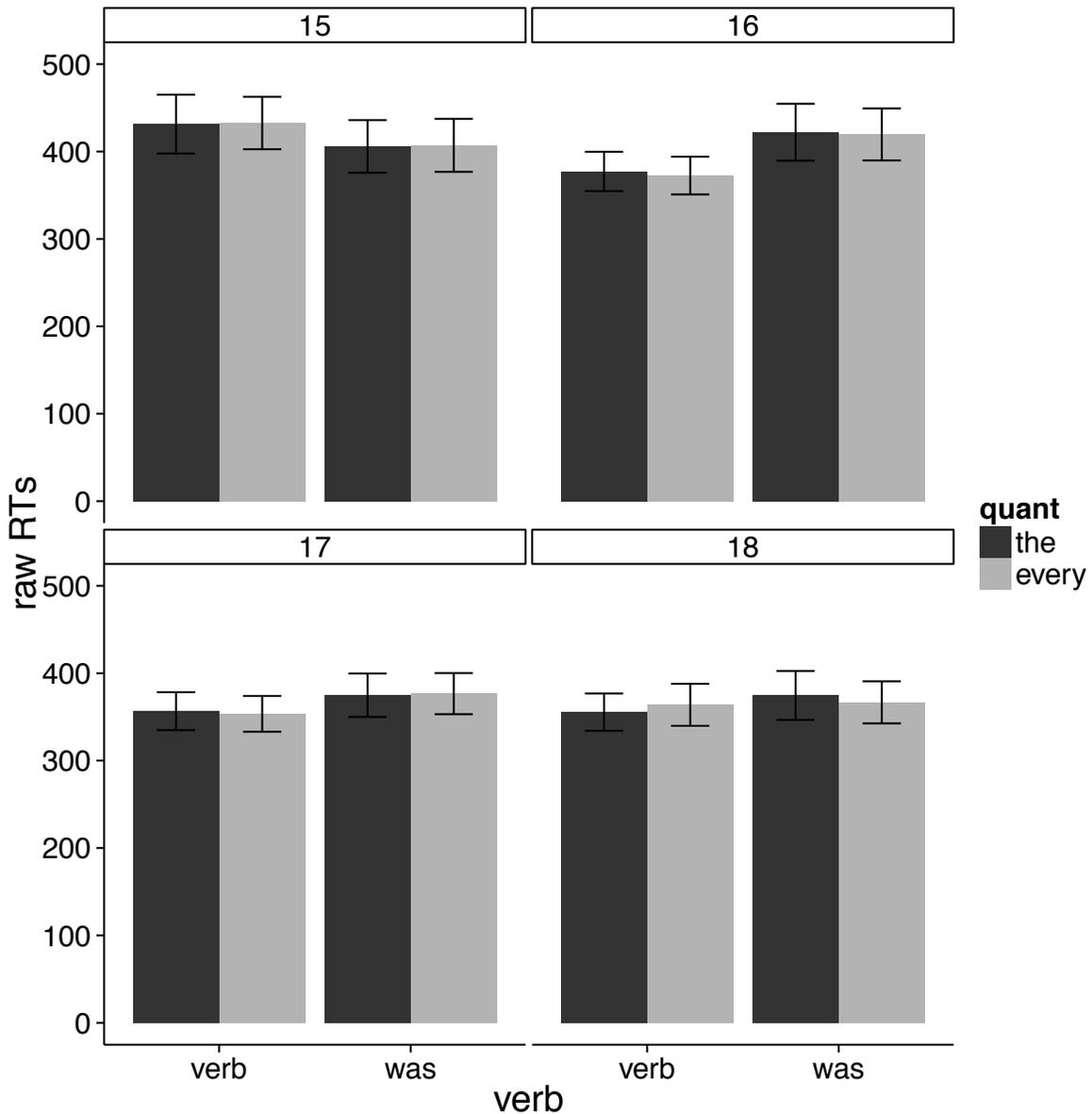

Figure 5: Raw reading times for HKV critical region in Replication of HKV Experiment 1. Error bars are 95% confidence intervals on participant means.

Following HKV, we analyze only correct trials in the analyses that we present here.

Following HKV, we performed participant and item ANOVAs on each of these regions. As HKV did, we found a significant main effect of verb type at word region 15 ("in"), such that the verb conditions were read more slowly than the ellipsis conditions (F1(1,71) = 7.81, p = .007; F2(1,57) = 24.79, p < .001). Furthermore, we also replicated HKV in finding that this effect reversed on words 16-18 ("order to arrange"), such that the ellipsis conditions were read more slowly than the verb conditions in each region (word 16: F1(1,71) = 33.82, p < .001; F2(1,57) = 55.05, p < .001; word 17: F1(1,71) = 17.99, p <

.001; F2(1,57) = 26.29, p < .001; word 18: F1(1,71) = 3.96, p = .05; F2(1,57) = 9.23, p = .004). But, unlike HKV, we did not find the predicted interaction between verb type and determiner in any of the four regions that we investigated (word 15: F1(1,71) = 0.046, p = .832; F2(1,57) = 0.113, p = .738; word 16: F1(1,71) = 0.011, p = .916; F2(1,57) = 0.023, p = .881; word 17: F1(1,71) = 0.083, p = .774; F2(1,57) = 0.006, p = .937; word 18: F1(1,71) = 1.30, p = .258; F2(1,57) = 8.88, p = .004). It should be noted that the predicted interaction was reliable in the items analyses at word 18, but this effect was not significant in the participants analysis. Furthermore, this effect is later than the effects that HKV report.

We also performed linear mixed-effect regressions in each of these regions, with the same conclusions. First, we replicated HKV's findings that (a) the verb conditions are read more slowly than the ellipsis conditions at word 15 (the word following the verb), and (b) the ellipsis conditions are read more slowly than the full verb conditions at the next three words, words 16-18. But we did not find the critical interaction between verb type and determiner in any of these regions. See the provided R analysis file for details (available at https://osf.io/t6anw ).

Finally, we entered HKV's residual RT data (after their data exclusions) together with our data (after our data exclusions) into a linear mixed effects regression at the critical word position as indicated by HKV, word position 17 (three words after the embedded verb). Here, we observed a main effect of verb type (beta = 19.83; t-value = 3.01), and a three-way interaction among verb-type, quantifier and Experiment (HKV vs. Gibson/Levy) (beta = 33.46; t = 2.443), and no other significant main effects or interactions. Thus there appears to be a significant difference between the results of the two experiments. We can investigate this a little further, by plotting the average size of the interaction effect for each participant in each experiment, and examining these distributions. The interaction effect was calculated by computing the difference of the differences associated with each quantifier (every, the) for each participant. This histogram is presented in Figure 6.

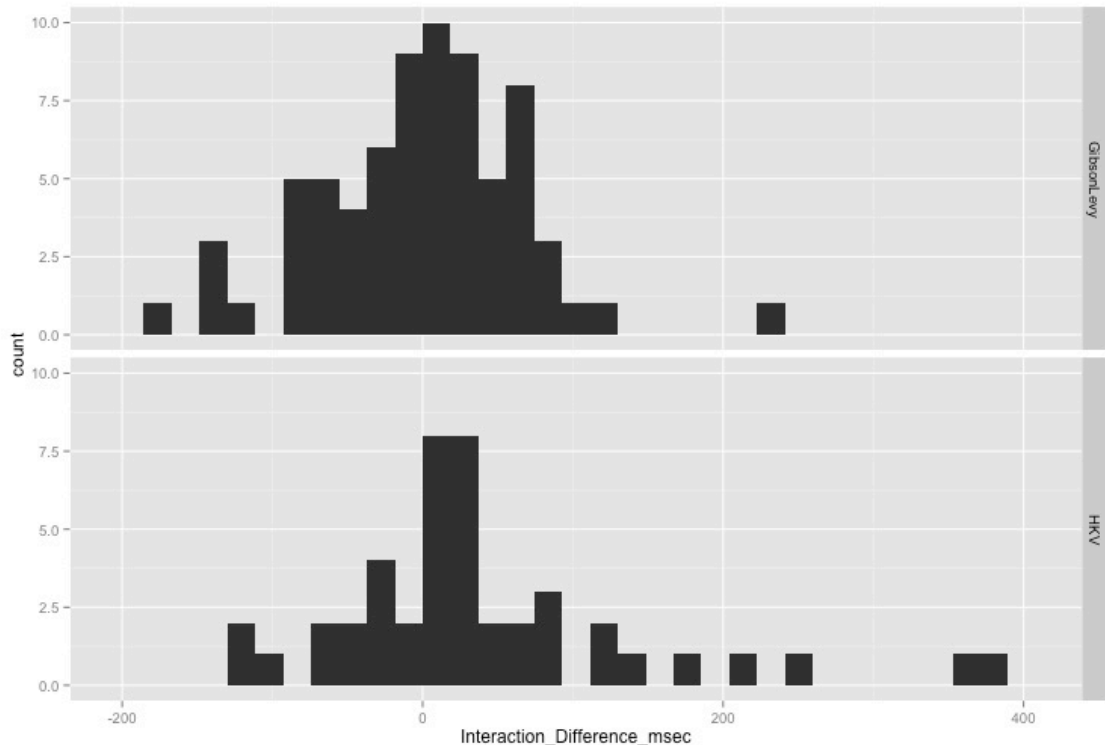

Figure 6: Histograms of the interaction effect for each participant in Experiment 1 in Gibson/Levy vs. comparable data from HKV, with data from trials answered correctly only. The distributions look very similar, with the exception that HKV's data have a few long RT outliers, which probably drive their observed effect.

We can see from the comparison of histograms in Figure 6 that the distributions look very similar, with the exception that HKV's data have a few long RT outliers, which probably drive their observed effect. Indeed, removing either of the outliers (subjects 26 or 32) causes the effect to no longer be significant. Critically, the distributions are centered in roughly the same place, with only slight differences in the outliers. HKV's hypothesis is that these data should be shifted to the right significantly, which doesn't appear to be the case, even for their own data.

## 3. Attempted replication of HKV Experiment 2

### 3.1 Procedure

We used the same in-house self-paced reading software written by Harry Tily to run HKV's Experiment 2 on Amazon.com's Mechanical Turk. Using this software, sentence materials were presented one word at a time, and reaction times between button presses were recorded. There was a comprehension question following each trial, as in HKV's self-paced reading experiments.

## 3.2 Materials

The target materials consisted of the original versions of HKV's 60 Experimental 2 items, as exemplified in (2):

(2) The doctor was reluctant to treat ...
    a. full-verb: the / every patient that the recently hired nurse admitted...
    b. local ellipsis: the / every patient that the recently hired nurse did...
    c. long-distance ellipsis: the / every patient that the recently hired nurse was...
    ... after looking over the test results.

At the time that we ran this replication attempt in 2012, we only had access to HKV's target materials, in an appendix in their paper, but we did not have access to HKV's filler materials or to the comprehension questions for the target materials. We therefore constructed our own set of 60 filler items, with questions, and comprehension questions for the target materials.

## 3.3 Participants

80 participants from Amazon.com's Mechanical Turk who did not take part in Experiment 1 took part in this self-paced reading study. We analyzed only participants who said that they were native speakers of English from the US. This resulted in omitting the data from 8 of the initial 80 participants. All participants remaining got at least 71% of the comprehension questions correct.

## 3.4 Results

Accuracy data. Participants answered 88.3% of the comprehension questions correctly, across target and filler materials, and 83.5% of the questions following only target items. The range of accuracies on particular target items varied from 50% to 100%, with two items that were miscoded, at 14% and 33%. We present analyses with items with at least 60% accuracy, thus excluding 8 of the 60 items (12, 16, 17, 22, 29, 40, 42, 60). The statistical analyses look very similar when all items are included.

Reading time data. As in Experiment 1, we used HKV's data-trimming procedures and analyzed residual RTs. We present the results from analyzing only correct trials, following HKV. Other ways of analyzing the data result in very similar statistical results. Following HKV, we focus our analysis on the four words following the disambiguating verb in the ACD relative clause: words 15-18 ("after looking over the" in (2)). Most critically for the QR hypothesis, HKV found a 3x2 interaction between the verb type (full-verb, local-ellipsis, long-distance ellipsis) and determiner (the, every) at the second word in this region. HKV also found a 2x2 interaction at this word position between the verb type and determiner, where the verb type was restricted to the full-verb and local-ellipsis conditions. HKV's main observation was that the "the" version was read more slowly than the corresponding "every" version only for the local-ellipsis conditions. In each of the other pair of conditions, the "every" condition was at least numerically slower than the "the" version.

We present means and 95% confidence intervals for each of the four words immediately following the verb region in the figures below, word regions 15-18 ("after looking over the" in (2)). Figure 7 presents the residual RTs across the sentences for Experiment 2. For comparison purposes, we also present HKV's Figure 2, on their residual RTs in their Experiment 1 in our Figure 8. Very similar RT patterns are visible across the two data sets. Figure 9 presents the residual RTs for our experiment in word regions 15-18. Then we present raw RTs across the sentences in Figure 10, and broken down by region in Figure 11.

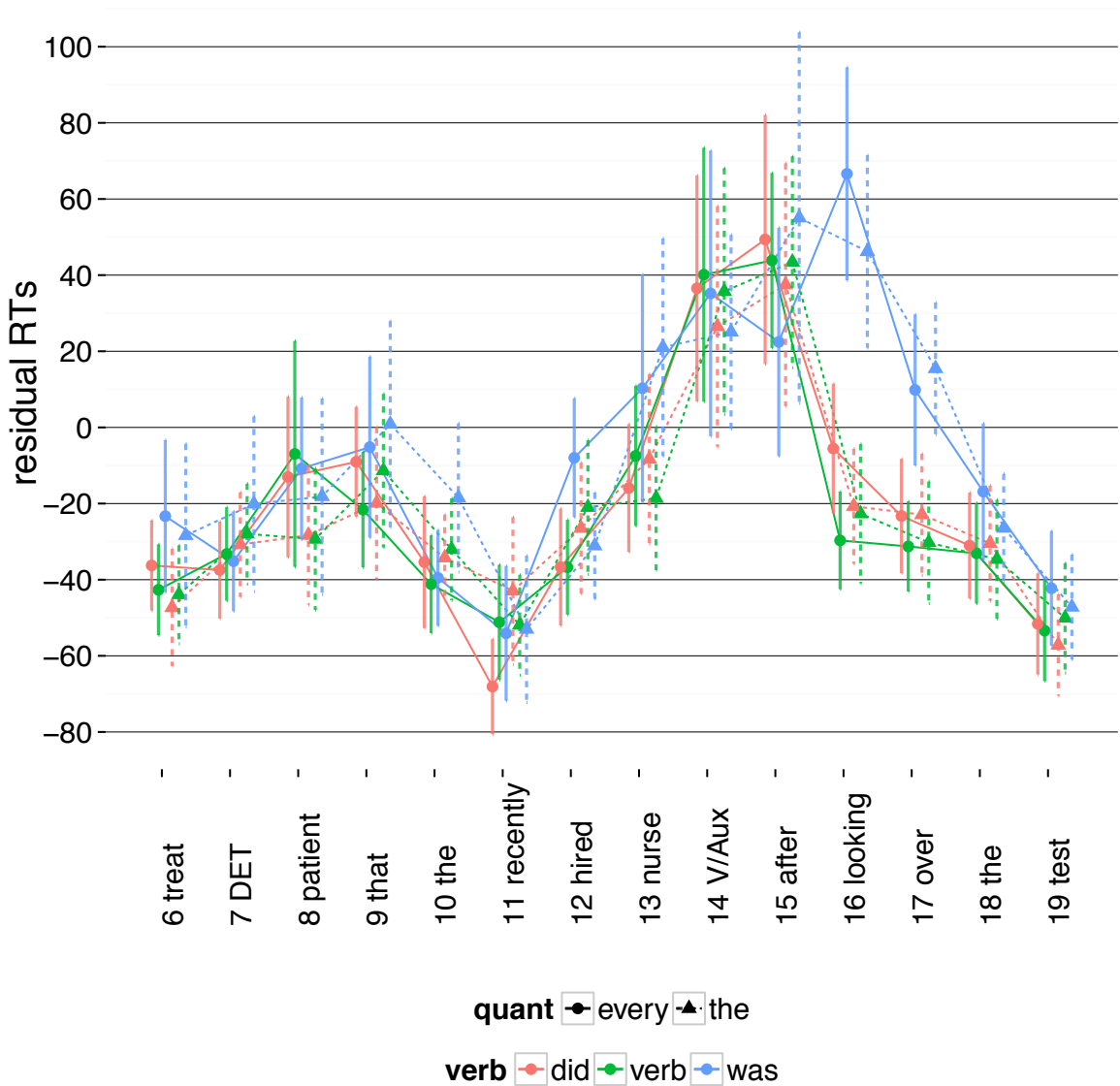

Figure 7: Word-by-word residual reading times for Replication of HKV Experiment 2. Error bars are 95% confidence intervals on participant means.

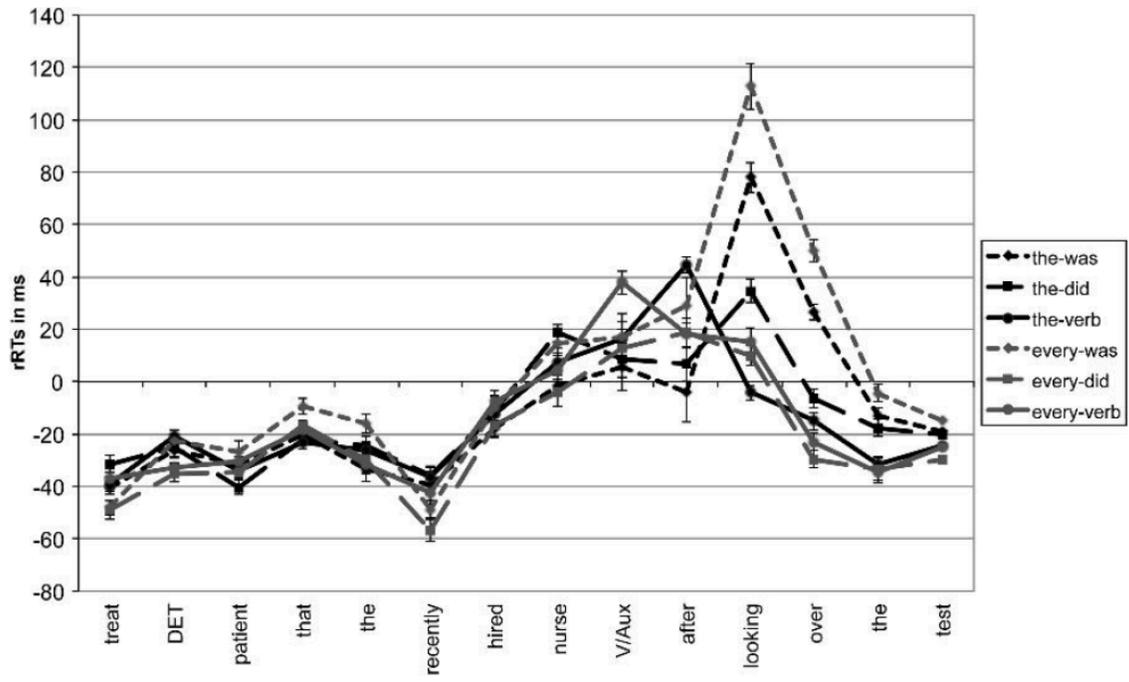

**Figure 4** ACD resolution by Determiner and Ellipsis-Size ($n = 48$)

Figure 8: Figure 4 from HKV: Average residual RTs for Experiment 2. Note that the error bars here are smaller than in our plots since they represent standard errors, as opposed to 95% CIs.

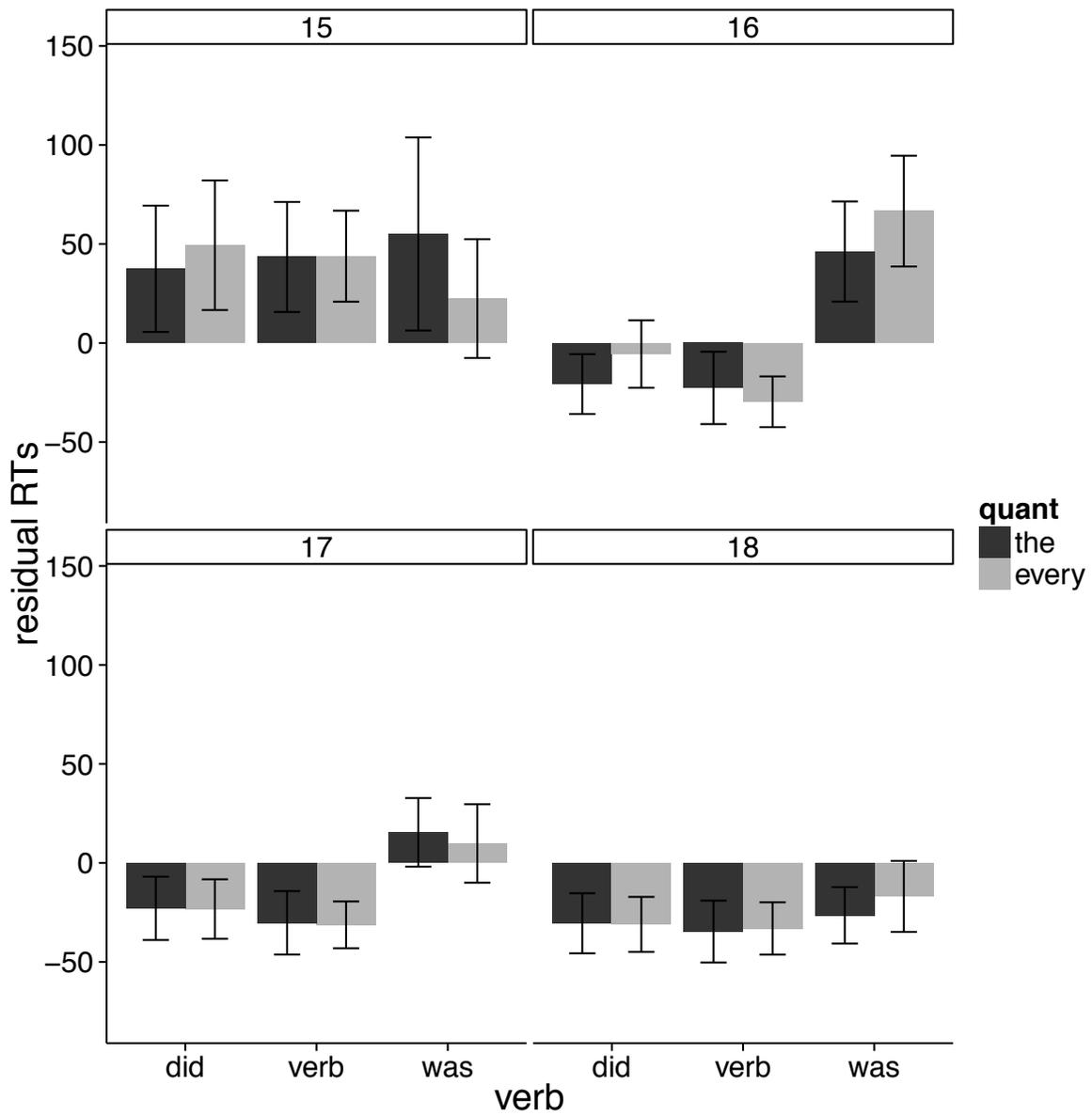

Figure 9: Critical region reading times for Replication of HKV Experiment 2. Error bars are 95% confidence intervals on participant means.

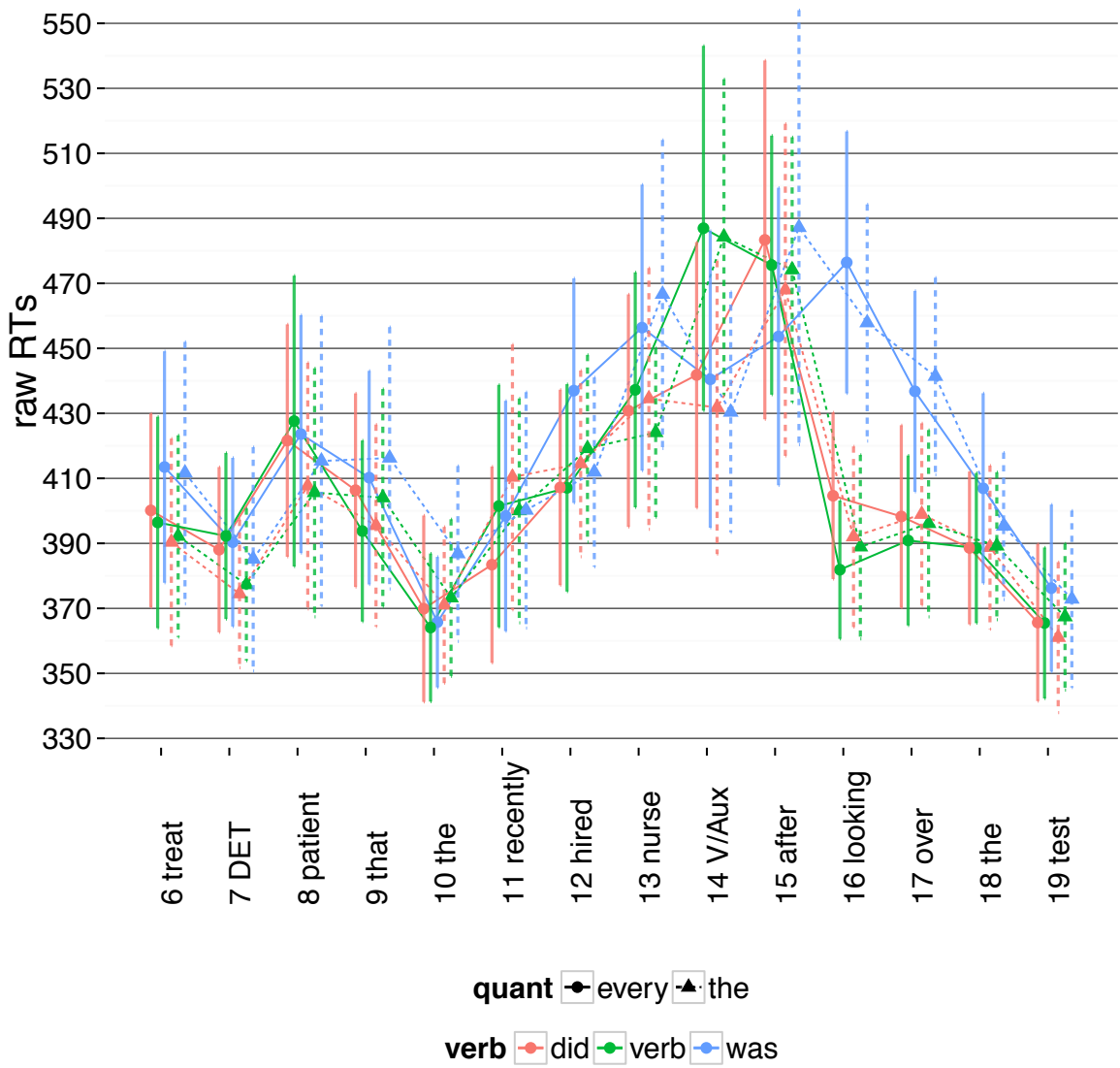

Figure 10: Word-by-word raw reading times for Replication of HKV Experiment 2. Error bars are 95% confidence intervals on participant means.

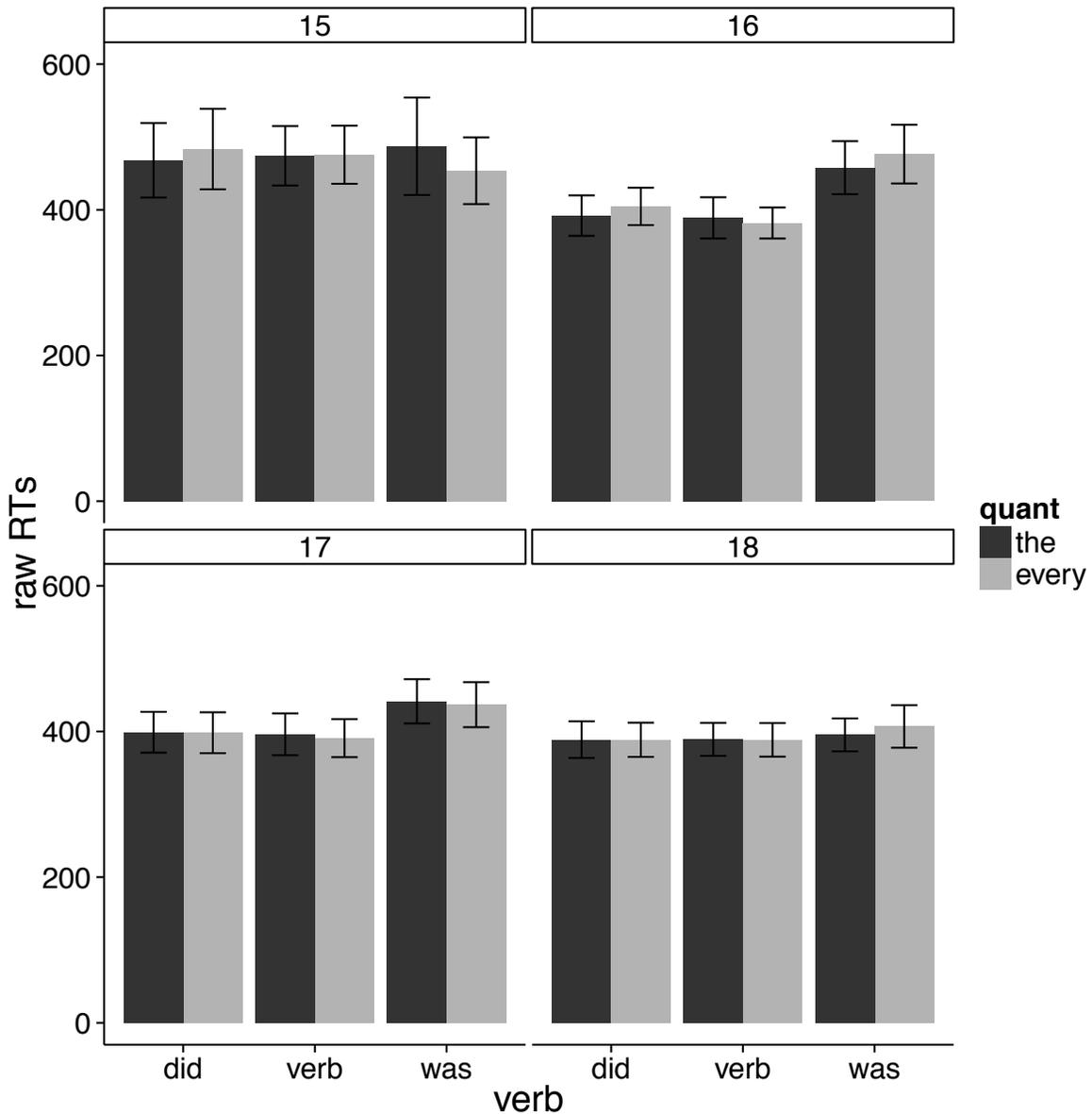

Figure 11: Raw reading times for Replication of HKV Experiment 2. Error bars are 95% confidence intervals on participant means.

Following HKV, we analyze only correct trials in the analyses that we present here, and we performed participant and item ANOVAs on each of these regions. As HKV did, we found a significant main effect of verb type at word regions 16 and 17 ("looking over" in (2)), such that the long-distance ellipsis conditions were read more slowly than the other conditions (word 16: $F1(2,144) = 36.84$, $p < .001$; $F2(2,102) = 40.16$, $p < .001$; word 17: $F1(2,144) = 28.32$, $p < .001$; $F2(2,102) = 4.17$, $p = .044$).

But, unlike HKV, we did not find the predicted interaction between verb type and determiner in the full 3x2 design: word 15: $F1(2,144) = 1.36$, $p = .26$; $F2(2,102) = 1.66$, $p$

= .195; word 16: F1(2,144) = 1.51, p = .225; F2(2,102) = 2.82, p = .065; word 17: F1(2,144) = 0.121, p = .886; F2(2,102) = 0.484, p = .618; word 18: F1(2,144) = 0.678, p = .509; F2(2,102) = 0.602, p = .55. Nor did we find the predicted interaction between verb type and determiner in the restricted 2x2 conditions, where the verb type is constrained to (full-verb, local-ellipsis): word 15: F1(1,72) = 0.307, p = .581; F2(1,51) = 0.955, p = .333; word 16: F1(1,72) = 2.478, p = .12; F2(1,51) = 3.53, p = .066; word 17: F1(1,72) = 0.005, p = .947; F2(1,51) = 0.422, p = .519; word 18: F1(1,72) = 0.067, p = .796; F2(1,51) = 0.843, p = .363. Note that although there was a marginally significant interaction at word position 16 in each of the 3x2 and 2x2 sets of conditions, this interaction is not the one that the QR predicts. In fact, this marginal interaction is actually the opposite of what is predicted by the QR hypothesis, such that the "every"-short-ellipsis condition is slower relative to the "the"-short-ellipsis condition.

Next we ran a mixed effect analysis. Because the maximal random effect structure did not converge, we instead ran models predicting residual reading time from determiner and verb type and their interaction, including random intercepts for subject and item, and random slopes by subject and item for the interaction term. These analyses showed no reliable interactions in any word position.

Finally, we entered HKV's residual RT data (after their data exclusions) together with our data (after our data exclusions) into a linear mixed effects regression at the critical word position as indicated by HKV, word position 16 (two words after the embedded verb). Here, we observed a three-way interaction among verb-type, quantifier and Experiment (HKV vs. Gibson/Levy) (beta = 47.00; t = -1.997), along with some other main effects and interactions. Thus there appears to be a significant difference between the results of the two experiments. We also investigated the two runs of the experiment, restricting our attention to short ellipsis conditions. Here, we again observed a three-way interaction among verb-type, quantifier and Experiment (HKV vs. Gibson/Levy) (beta = -65.36; t = -3.758), along with some other main effects and interactions. Thus there appears to be a significant difference between the results of the two experiments.

We can investigate this further by plotting the average size of the interaction effect for each participant in each experiment, and examining these distributions. The interaction effect was calculated by computing the difference of the differences associated with each quantifier (every, the) and the two short ellipsis conditions for each participant. This histogram is presented in Figure 12.

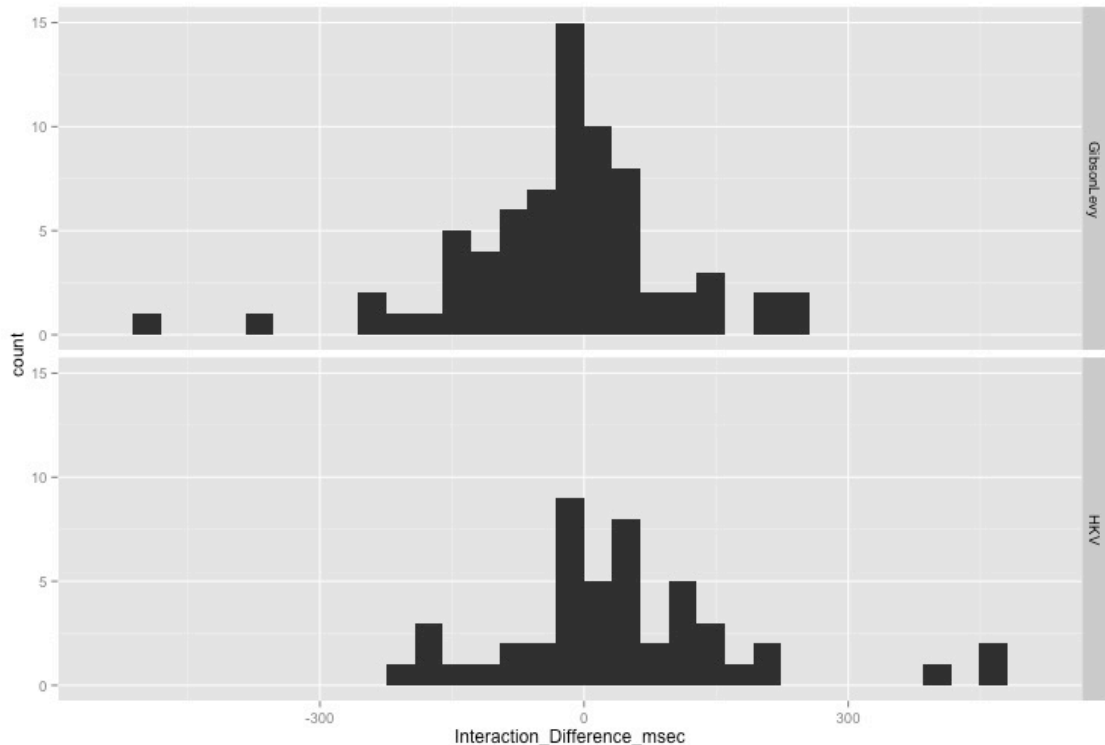

Figure 12: Histograms of the interaction effect for each participant in Experiment 2 in Gibson/Levy vs. comparable data from HKV, with data from trials answered correctly only. The distributions look very similar, with the exception that HKV's data have a few long RT outliers, which probably drive their observed effect.

We can see from the comparison of histograms in Figure 6 that the distributions look very similar, with the exception that HKV's data have a few long RT outliers, which probably drive their observed effect. Indeed, removing either of the outliers (subjects 3 or 10) causes the effect to no longer be significant. The Gibson/Levy data also have a few short outliers at this region, contributing to the strength of the difference between the two experimental runs. But critically, the distributions are centered in roughly the same place, with the only important differences being in the outliers. HKV's hypothesis is that these data should be shifted to the right significantly, which doesn't appear to be the case, even for their own data.

## 4. Summary and concluding remarks

HKV reported reading time evidence that purported to provide evidence for the quantifier-raising hypothesis. The evidence consisted of an interaction in the second or third word following the embedded relative clause verb in examples like (1) and (2), such that the "the" / local-ellipsis condition was read especially slowly compared to its controls. In this paper, we have attempted to replicate both of HKV's self-paced reading experiments, but we have failed to find even hints of the predicted interactions in the locations that HKV found them.

In other work, we have analyzed HKV's reading time data (obtained from HKV), and we have discovered several unreported choice points, errors, and concerns regarding multiple comparisons in the original HKV data analysis (Gibson et al., submitted). Importantly, there is no analysis that produces a significant interaction in both experiments if one analyzes the same region across experiments: HKV analyze the word 3 words after the relative clause verb in Experiment 1, but they analyze the word 2 words after the relative clause verb in Experiment 2. Furthermore, analyses with all trials – not just trials with correct answers to comprehension questions – or analyses with the additional conditions that were run in Experiment 1, or analyses with different ways of removing low-accuracy participants and items – result in the crucial interaction being non-significant.

Putting these observation together with the failure to observe the crucial interaction in either of the experiments that we report here, we conclude that the experiments reported by HKV should not be viewed as providing evidence for the ACD quantifier raising processing effect.